\documentclass[a4paper,12pt]{article}

\usepackage{amsmath,amssymb}
\usepackage{bbm}
\usepackage{slashed}
\usepackage{feynmp}
\usepackage[nosort]{cite}
\usepackage{graphicx}
\usepackage[bookmarks=true]{hyperref}

\usepackage{ifpdf}
\ifpdf
\fi

\ifx\hypersetup\sadfkjashdfkxja\else
\hypersetup{pdftitle={}}
\hypersetup{pdfauthor={}}
\hypersetup{pdfsubject={}}
\hypersetup{pdfkeywords={}}
\hypersetup{plainpages=false}
\hypersetup{bookmarksnumbered=true}
\hypersetup{pdfstartview=FitH}
\hypersetup{pdfpagemode=UseNone}
\hypersetup{colorlinks=false}
\hypersetup{citebordercolor={.5 1 .5}}
\hypersetup{urlbordercolor={.5 1 1}}
\hypersetup{linkbordercolor={1 .7 .7}}
\fi

\newcommand{\Tr}{\mathop{\mathrm{Tr}}}

\def\be{\begin{equation}}
\def\ee{\end{equation}}
\def\bsp{\be\begin{split}}
\def\la{\langle}
\def\ra{\rangle}
\def\dag{\dagger}

\def\G{\Gamma}
\def\D{\Delta}

\def\a{\alpha}
\def\b{\beta}
\def\g{\gamma}

\def\d{\delta}
\def\e{\epsilon}
\def\m{\mu}

\def\n{\nu}

\def\l{\lambda}

\def\o{\omega}

\def\p{\partial}

\renewcommand{\title}[1]{\vbox{\center\LARGE{#1}}\vspace{5mm}}
\renewcommand{\author}[1]{\vbox{\center\large{#1}}\vspace{5mm}}
\newcommand{\address}[1]{\vbox{\center\em#1}}
\newcommand{\email}[1]{\vbox{\center\tt#1}\vspace{5mm}}

\begin{document}
\begin{fmffile}{c3po1}
\bibliographystyle{utphys}

\begin{titlepage}
\hfill {\tt QMUL-PH-14-10}\\
\title{\vspace{1.0in} {\bf ABJ(M) Chiral Primary Three-Point Function at
    Two-loops}}
 
\author{Donovan Young}

\address{Centre for Research in String Theory\\
School of Physics and Astronomy\\
Queen Mary, University of London\\
Mile End Road, London E1 4NS, United Kingdom}

\email{d.young@qmul.ac.uk}

\abstract{We compute the leading correction to the structure constant
  for the three-point function of three length-two chiral primary
  operators in planar ABJ(M) theory at weak 't Hooft coupling.}

\end{titlepage}

\section{Introduction}

The AdS$_4$/CFT$_3$ correspondence between the ${\cal N}=6$
superconformal Chern-Simons-matter theory of Aharony, Bergman,
Jafferis, and Maldacena (ABJM) \cite{Aharony:2008ug} and M-theory on
$AdS_4\times S^7/\mathbb{Z}_k$ presents a fertile playground for
further explorations of the gauge-gravity duality, beyond the
well-trodden ground of the AdS$_5$/CFT$_4$ correspondence between
${\cal N}=4$ supersymmetric Yang-Mills in four dimensions and type-IIB
strings on $AdS_5\times S^5$. In many respects one expects a very
similar picture to that found so far for ${\cal N}=4$ SYM: planar
integrability, localization formulae, supersymmetric Wilson loops,
scattering amplitudes, etc. have been found on both sides of this
AdS$_4$/CFT$_3$ correspondence. In many cases ABJM simply presents an
added complication, as in the case of perturbation theory where often
the analogue to one-loop corrections in ${\cal N}=4$ SYM are two-loop
corrections in ABJM. 

A point of departure is found however, in the three-point correlation
functions of chiral primary operators (CPO's) of dimension $J$
\be
{\cal O}^J_{A} =\frac{1}{\sqrt{J/2}}\left(\frac{k}{4\pi\sqrt{NM}}\right)^{J/2}
 ({\cal C}_{A})^{A_1\ldots A_{J/2}}_{B_1\ldots B_{J/2}} \,
\Tr \left(Y^{B_1} Y^\dag_{A_1} \cdots Y^{B_{J/2}} Y^\dag_{A_{J/2}}\right),
\ee
built from the $U(N)\times U(M)$ bifundamental scalar fields $Y^A$ of
the ABJ(M) \cite{Aharony:2008gk} theory using the tensors ${\cal C}_A$
which are symmetric in upper and in lower indices and are traceless in
any pair consisting of one upper and one lower index\footnote{In this
  normalization the two-point function is given by $\la{\cal
    O}_A^J(x){\cal O}_B^J(0)\ra = \d_{AB}/(4\pi |x|)^J $.}. These
operators have vanishing anomalous dimension. Unlike in ${\cal N}=4$
SYM, where chiral primary operators have protected three-point
functions (c.f. \cite{Baggio:2012rr} and references therein), in ABJM
we know from supergravity \cite{Bastianelli:1999en} that they scale as
$\l^{1/4}/N$ where $\l =N/k$ is the 't Hooft coupling\footnote{ABJM
  implies that $N=M$. When discussing results where $N$ and $M$ are
  distinct we will use $\l = N/k$ and $\hat\l = M/k$.}. Given the
conformally-fixed form
\be
\bigl\la {\cal O}_1(x_1) {\cal O}_2(x_2) {\cal O}_3(x_3) \bigr\ra = 
\frac{1}{(4\pi)^\g}
\frac{C_{123}(\l)}
{|x_{12}|^{\g_3} |x_{23}|^{\g_1} |x_{31}|^{\g_2}},
\ee
where $\gamma_i = (\sum_jJ_j-2J_i)/2$, $\g=\g_1+\g_2+\g_3$, and
$x_{ij}=x_i-x_j$, the specific pattern of structure constants at
strong coupling and for large $N=M$ was given in \cite{Hirano:2012vz}
(we take $J_3\geq J_2 \geq J_1$)
\bsp\label{sc}
&C_{123}(\l\gg 1) =\\
&\frac{1}{N} \left(\frac{\l}{2\pi^2}\right)^{1/4}
\frac{\prod_{i=1}^3 
\sqrt{J_i+1}\,(J_i/2)!\,\G(\g_i/2+1) }{\Gamma(\g/2+1)} \\
&\sum_{p=0}^{\g_3}
\frac{\left({\cal C}_1\right)^{I_1\ldots I_p I_{p+1}\ldots I_{J_1/2}}
_{K_1 \ldots K_{\g_3 -p} K_{\g_3-p+1} \ldots K_{J_1/2}}
\left({\cal C}_2\right)^{K_1 \ldots K_{\g_3 -p} L_1\ldots
L_{\g_1-J_2/2+p}}_{I_1\ldots I_p M_1 \ldots M_{J_2/2-p}}
\left({\cal C}_3\right)^{K_{\g_3-p+1} \ldots K_{J_1/2}M_1 \ldots
  M_{J_2/2-p}}
_{I_{p+1}\ldots I_{J_1/2}  L_1\ldots L_{\g_1-J_2/2+p}}}
{p! (\g_3-p)! (\g_1-J_2/2+p)!
(J_2/2-p)!(\g_2-J_1/2+p)! (J_1/2-p)!},
\end{split}
\ee 
in terms of the three tensors ${\cal C}$ defining the three
CPO's. This is a remarkable difference from ${\cal N}=4$ SYM: not only
do the structure constants $C_{123}$ depend on the coupling $\l$, they
consist of a host of interpolating functions -- one for each $\g_3+1$
ways\footnote{\label{foot:s}Switching all upper and lower indices simultaneously on
  all ${\cal C}$ tensors is a symmetry of the three-point functions.
  This reduces the number $\g_3+1 \to (\g_3+1)/2$ if $\g_3$ is odd or
  $(\g_3+2)/2$ otherwise.} the ${\cal C}$ tensors can be
contracted. These interpolating functions depend on the dimensions of
the operators $J_i$ in a non-trivial way. Why should one associate an
interpolating function to each possible ${\cal C}$ tensor contraction?
Because at tree-level in the planar theory there is a single
way\footnote{\label{foot:t} This statement is true when the $\g_i$ are
  even, which means that each ${\cal C}$ tensor shares as many up as
  down indices with each other ${\cal C}$ tensor. In this case
  $C_{123}$ carries a colour factor of $C_F=(1/N+1/M)$. In the case of odd
  $\g_i$ there are two contractions of the ${\cal C}$ tensors which
  are related by switching all down and up indices simultaneously on
  all ${\cal C}$ tensors. In this case the colour factor is
  $C_F=1/\sqrt{NM}$.} in which the ${\cal C}$ tensors contract, which is
included in the sum over $p$ above and which we will label as $\la
{\cal C}_1\,{\cal C}_2\,{\cal C}_3\ra_{\text{tree}}$ (the colour
factor $C_F$ is given in footnote \ref{foot:t})
\be
C_{123}(\l\ll 1) = C_F
\sqrt{(J_1/2)(J_2/2)(J_3/2)}\, \la {\cal C}_1\,{\cal
  C}_2\,{\cal C}_3\ra_{\text{tree}} + \text{loop corrections}.
\ee
One therefore sees that the various interpolating functions kick-in at
higher orders as the 't Hooft coupling is increased. This is not true
for the extremal correlators where $J_3=J_1+J_2$ -- they retain this
form even at strong coupling because $p$ is forced to zero
\be
C_{123}^{\text{extremal}}(\l\gg 1)=
\frac{1}{N} \left(\frac{\l}{2\pi^2}\right)^{1/4}
\sqrt{(J_1+1)(J_2+1)(J_3+1)}\, \la {\cal C}_{1}{\cal
  C}_{2}{\cal C}_{3}\ra_{\text{tree}}.
\ee

\begin{figure}
\[\int d^{2\o}x_1~
\parbox{20mm}{
\begin{fmfgraph*}(50,50)
\fmftop{v1}
\fmfbottom{v2,v3}
\fmf{plain,right=0.5}{v3,va}
\fmf{plain,left=0.5}{v3,va}
\fmf{plain,right=0.5}{v2,va}
\fmf{plain,left=0.5}{v2,va}
\fmf{plain,right=0.5}{v1,va}
\fmf{plain,left=0.5}{v1,va}
\fmfv{decor.shape=circle,decor.filled=30,decor.size=9}{v1}
\fmfv{decor.shape=circle,decor.filled=30,decor.size=9}{v2}
\fmfv{decor.shape=circle,decor.filled=30,decor.size=9}{v3}
\fmfv{d.sh=circle,l.d=0, d.f=10,d.si=.6w,l=$2$}{va}
\end{fmfgraph*}}
~~~=~~~
\parbox{20mm}{
\begin{fmfgraph*}(50,50)
\fmftop{v1}
\fmfbottom{v2,v3}
\fmf{plain,right=0.35}{v3,va}
\fmf{plain,left=0.35}{v3,va}
\fmf{plain,right=0.35}{v2,va}
\fmf{plain,left=0.35}{v2,va}
\fmf{dbl_plain,right=1}{v1,va}
\fmf{dbl_plain,left=1}{v1,va}
\fmfv{decor.shape=circle,decor.filled=30,decor.size=9}{v2}
\fmfv{decor.shape=circle,decor.filled=30,decor.size=9}{v3}
\fmfv{d.sh=circle,l.d=0, d.f=10,d.si=.6w,l=$2$}{va}
\end{fmfgraph*}}\]
\caption{The main idea behind the method: two-loop corrections to the
  three-point function are gotten via integration over an external
  point thus transforming three-point integrals into three-loop
  propagator-type integrals. The line associated with the integrated
  operator has a doubled propagator, i.e. $1/p^4$ in momentum space.}
\label{fig:main}
\end{figure}
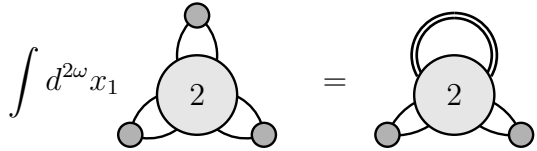

In this paper we will take the first steps towards exploring the
$C_{123}(\l)$ in perturbation theory. We will focus on one of the
aforementioned interpolating functions arising from three length-two
operators. In this setting there are just two possible ways of
contracting the ${\cal C}$ tensors, but we will choose operators such
that only one is non-zero\footnote{In any case, by footnote
  \ref{foot:s} there is only one interpolating function at play.}. The
first correction appears at ${\cal O}(\l^2)$ or two-loops. This
presents a challenge because three-loop integrals with three off-shell
legs are difficult to work with and have not been widely considered in
the literature. In order to overcome this obstacle, we integrate in
configuration space over one of the operators' position using
dimensional regularization, see figure \ref{fig:main}. This reduces
the problem to three-loop propagator diagrams which have been widely
studied and are tractable. We believe that this trick works when
dealing with protected operators since the three-point function is
guaranteed to be finite {\it and} the coordinate dependence is fixed
to a known form by conformal symmetry, where the powers of the
coordinate differences are independent of the coupling.

We begin in section \ref{sec:method} with a presentation of the
details of our method. In section \ref{sec:n4} we give an exhibition
of it in the setting of ${\cal N}=4$ SYM at the one-loop level where
we show that the three-point function indeed comes out uncorrected,
i.e. independent of the coupling $g_{YM}^2N$. We continue in section
\ref{sec:main} with our main result, the structure constant for three
length-two CPO's in ABJ(M), given by (\ref{bres}). The calculational
method is that employed originally in \cite{Minahan:2009wg} and
recently in an almost identical setting (two-loop form factors for the
same operators) in \cite{Young:2013hda}. Finally we end with a
discussion in section \ref{sec:disc} and give details of the
calculation in two appendices.

\section{Method}
\label{sec:method}

We consider the three CPO's
\be
{\cal O}_1 = \frac{k}{4\pi\sqrt{MN}}\Tr( Y^1Y^\dag_2),~~
{\cal O}_2 = \frac{k}{4\pi\sqrt{MN}}\Tr( Y^2Y^\dag_3),~~
{\cal O}_3 = \frac{k}{4\pi\sqrt{MN}}\Tr( Y^3Y^\dag_1).
\ee
We know from conformal invariance that\footnote{From now on we choose
  to factor the colour factor out of the structure constant.}
\be
\bigl\la {\cal O}_1(x_1) {\cal O}_2(x_2) {\cal O}_3(x_3) \bigr\ra =
\frac{1}{(4\pi)^3\sqrt{NM}}\frac{\hat C_{123}(\l)}
{|x_{12}| |x_{23}| |x_{31}|},
\ee
where $\hat C_{123} =1 + c_1\l^2+c_2\hat\l^2+c_3\l\hat\l+\ldots$.
We continue this expression to $2\o = d-2\e$ dimensions
\be\label{threeptdef}
\bigl\la {\cal O}_1(x_1) {\cal O}_2(x_2) {\cal O}_3(x_3) \bigr\ra =
\frac{1}{\sqrt{NM}}
\frac{\hat C_{123}(\l)}
{\left(x^2_{12} x^2_{23} x^2_{31}\right)^{\o-1}} \,\frac{\G^3(\o-1)}{(4\pi^\o)^3},
\ee
and integrate over $x_1$ using dimensional regularization. We obtain
\be\label{threeptint}
\sqrt{MN}\int d^{2\o} x_1 \,\la {\cal O}_1(x_1) {\cal O}_2(x_2) {\cal O}_3(x_3)
\ra 
= \frac{\G(\o-1)\G(\o-2)}{2^6\pi^{2\o}(x_{23}^2)^{2\o-3}}\, \hat C_{123}(\l).
\ee
This integration is free from UV divergences in any dimension. It is
however IR divergent in any dimension $\leq 4$. Thus the method leaves
unmolested the short distance physics determining the renormalization
of the operators, but does introduce IR divergences into the loop
integrations in momentum space. Since the method relies crucially on
using one and the same renormalization parameter to regulate these two
classes of divergences one may encounter the accidental cancellation
of IR and UV poles in intermediate stages in the calculation, as we
will see for example in section \ref{sec:n4}, see footnote
\ref{foot:b}. However, this is not a cause for concern: the RHS of
(\ref{threeptdef}) is a guaranteed finite quantity. Thus the only true
divergences are the IR divergences introduced by the integration over
$x_1$ and these are regulated in a controlled way, i.e. by the RHS of
(\ref{threeptint}).

Using (\ref{threeptint}) we may express $\hat C_{123}(\l)$ via
\be\label{C}
\hat C_{123}(\l)=
\sqrt{NM}\frac{2^6\pi^{2\o}(x_{23}^2)^{2\o-3}}{\G(\o-1)\G(\o-2)}
\int d^{2\o} x_1 \,\bigl\la {\cal O}_1(x_1) {\cal O}_2(x_2) {\cal O}_3(x_3)
\bigr\ra. 
\ee
In general $\hat C_{123}(\l)$ will be renormalized by the two-point
function
\be\label{g}
\bigl\la {\cal O}_i(x) {\cal O}_i(0) \bigr\ra =
\frac{\G^2(\o-1)}{(4\pi^\o)^2}
\frac{g_i(\l)}{(x^2)^{2(\o-1)}},
\ee
where $g_i=g = 1 + d_1\l^2+d_2\hat\l^2+d_3\l\hat\l+\ldots$, giving (c.f. \cite{Georgiou:2012zj})
\be
C_{123}\bigl|_{{\cal O}(\l^2)} = \left[\hat C_{123}
-\frac{1}{2}\sum_{i=1}^3 g_i
\right]_{{\cal O}(\l^2)}.
\ee
%

\section{${\cal N}=4$ SYM at one loop}
\label{sec:n4}

Before moving on to ABJM, it is instructive to see how the method
described in section \ref{sec:method} works in the case of ${\cal
  N}=4$ SYM at one-loop, where for CPO's we expect to find that there
is no correction to $C_{123}$. In analogy with the ABJM case we take
three length-two CPO's (here $\l = g_{YM}^2N$, where $g_{YM}$ is the Yang-Mills
coupling constant)
\be
{\cal O}_1 = \frac{1}{\sqrt{\l}}\Tr(\Phi^1\Phi^2),\quad
{\cal O}_2 = \frac{1}{\sqrt{\l}}\Tr(\Phi^2\Phi^3),\quad
{\cal O}_3 = \frac{1}{\sqrt{\l}}\Tr(\Phi^3\Phi^1),\quad
\ee
which we represent by three grey blobs in the diagrams below. Here we
use dimensional reduction in order to preserve supersymmetry. At one
loop we find that the contributions to the three-point function are as
follows, where we employ a double line to denote a propagator with
doubled weight, i.e. $1/p^4$ instead of $1/p^2$. Some details of the
calculation are provided in appendix \ref{app:n4}\footnote{\label{foot:b}In the
  fourth line below we see that the bubble at the top of the first
  triangle integrates to zero. This is an example of the
  coincidental UV-IR pole cancellation discussed in the previous section.}.

\[
\hspace{-5cm}
\int d^{2\o}x_1
\parbox{20mm}{
\begin{fmfgraph*}(50,50)
\fmftop{v1}
\fmfbottom{v2,v3}
\fmf{plain}{v1,va,v2}
\fmf{plain}{v1,vb,v3}
\fmf{plain}{v2,v3}
\fmffreeze
\fmf{wiggly}{va,vb}
\fmfv{decor.shape=circle,decor.filled=30,decor.size=9}{v1}
\fmfv{decor.shape=circle,decor.filled=30,decor.size=9}{v2}
\fmfv{decor.shape=circle,decor.filled=30,decor.size=9}{v3}
\end{fmfgraph*}}= 2 ~
\parbox{20mm}{
\begin{fmfgraph*}(60,60)
\fmfleft{i}
\fmfright{o}
\fmftop{t}
\fmf{plain,tension=1.4}{i,v1}
\fmf{plain,tension=1.4}{v2,o}
\fmf{plain, right,tension=1}{v1,v2}
\fmffreeze
\fmf{phantom, tension=.8}{t,v3}
\fmf{plain, right=0.4,tension=0.54}{v3,v1}
\fmf{plain, right=0.4,tension=0.54}{v2,v3}
\fmfv{d.sh=circle,l.d=0, d.f=empty,d.si=.2w}{v3}
\end{fmfgraph*}}
~+2~
\parbox{20mm}{
\begin{fmfgraph*}(60,60)
\fmfleft{i}
\fmfright{o}
\fmftop{t}
\fmf{plain,tension=1.4}{i,v1}
\fmf{plain,tension=1.4}{v2,o}
\fmf{plain, right,tension=1}{v1,v2}
\fmffreeze
\fmf{phantom, tension=.8}{t,v3}
\fmf{dbl_plain, right=0.4,tension=0.4}{v3,v1}
\fmf{plain, left=0.64,tension=0.4}{v3,v1}
\fmf{plain, right=0.4,tension=0.4}{v2,v3}
\end{fmfgraph*}}
\]
\vspace{-0.cm}
\[
\hspace{-1.25cm}
\int d^{2\o}x_1 \Biggl(\quad
\parbox{20mm}{
\begin{fmfgraph*}(50,50)
\fmftop{v1}
\fmfbottom{v2,v3}
\fmf{plain}{v1,va,v2}
\fmf{plain}{v1,v3}
\fmf{plain}{v2,vb,v3}
\fmffreeze
\fmf{wiggly}{va,vb}
\fmfv{decor.shape=circle,decor.filled=30,decor.size=9}{v1}
\fmfv{decor.shape=circle,decor.filled=30,decor.size=9}{v2}
\fmfv{decor.shape=circle,decor.filled=30,decor.size=9}{v3}
\end{fmfgraph*}}
+~
\parbox{20mm}{
\begin{fmfgraph*}(50,50)
\fmftop{v1}
\fmfbottom{v2,v3}
\fmf{plain}{v1,v2}
\fmf{plain}{v1,va,v3}
\fmf{plain}{v2,vb,v3}
\fmffreeze
\fmf{wiggly}{va,vb}
\fmfv{decor.shape=circle,decor.filled=30,decor.size=9}{v1}
\fmfv{decor.shape=circle,decor.filled=30,decor.size=9}{v2}
\fmfv{decor.shape=circle,decor.filled=30,decor.size=9}{v3}
\end{fmfgraph*}}
~\Biggr) =2\Biggl(
\parbox{20mm}{
\begin{fmfgraph*}(60,60)
\fmfleft{i}
\fmfright{o}
\fmftop{t}
\fmf{plain,tension=1.4}{i,v1}
\fmf{plain,tension=1.4}{v2,o}
\fmf{plain, right,tension=1}{v1,v2}
\fmffreeze
\fmf{phantom, tension=.8}{t,v3}
\fmf{plain, right=0.4,tension=0.4}{v3,v1}
\fmf{plain, left=0.64,tension=0.4}{v3,v1}
\fmf{dbl_plain, right=0.4,tension=0.4}{v2,v3}
\end{fmfgraph*}}
~+
\parbox{20mm}{
\begin{fmfgraph*}(60,60)
\fmfleft{i}
\fmfright{o}
\fmftop{t}
\fmf{plain,tension=1.4}{i,v1}
\fmf{plain,tension=1.4}{v2,o}
\fmf{plain, right,tension=1}{v1,v2}
\fmffreeze
\fmf{phantom, tension=.8}{t,v3}
\fmf{dbl_plain, right=0.4,tension=0.4}{v3,v1}
\fmf{plain, left=0.64,tension=0.4}{v3,v1}
\fmf{plain, right=0.4,tension=0.4}{v2,v3}
\end{fmfgraph*}}
\]
\vspace{-0.5cm}
\[+~
\parbox{20mm}{
\begin{fmfgraph*}(60,60)
\fmfleft{i}
\fmfright{o}
\fmftop{t}
\fmf{plain,tension=1.4}{i,v1}
\fmf{plain,tension=1.4}{v2,o}
\fmf{dbl_plain, right,tension=1}{v1,v2}
\fmffreeze
\fmf{phantom, tension=.8}{t,v3}
\fmf{plain, right=0.4,tension=0.4}{v3,v1}
\fmf{plain, left=0.64,tension=0.4}{v3,v1}
\fmf{plain, right=0.4,tension=0.4}{v2,v3}
\end{fmfgraph*}}
~ +
\parbox{20mm}{
\begin{fmfgraph*}(60,60)
\fmfleft{i}
\fmfright{o}
\fmf{plain, tension=2}{i,v1}
\fmf{plain, tension=2}{v3,o}
\fmf{plain,left=0.4, tension=1}{v1,v4,v3,v2,v1}
\fmffixed{(0,26)}{v2,v4}
\fmf{plain}{v2,v4}
\end{fmfgraph*}}
~ -2p^2~
\parbox{20mm}{
\begin{fmfgraph*}(50,50)
\fmfleft{i}
\fmfright{o}
\fmf{plain, tension=2}{i,v1}
\fmf{plain, tension=2}{v3,o}
\fmf{plain,left=0.4, tension=1}{v1,v4,v3,v2,v1}
\fmffixed{(0,26)}{v2,v4}
\fmf{plain}{v2,v4}
\fmffreeze
\fmf{dbl_plain,left=0.4, tension=1}{v1,v4}
\end{fmfgraph*}}
-~
\parbox{20mm}{
\begin{fmfgraph*}(50,50)
\fmfleft{i}
\fmfright{o}
\fmf{plain,tension=5.4}{i,v1}
\fmf{plain,tension=5.4}{v2,o}
\fmf{plain, right,tension=1}{v1,va}
\fmf{plain, left,tension=1}{v1,va}
\fmf{dbl_plain, right,tension=1}{v2,va}
\fmf{plain, left,tension=1}{v2,va}
\end{fmfgraph*}}
\Biggr)
\]
\vspace{-0.cm}
\[
\hspace{-1.75cm}
\int d^{2\o}x_1 \Biggl(\quad
\parbox{20mm}{
\begin{fmfgraph*}(50,50)
\fmftop{v1}
\fmfbottom{v2,v3}
\fmf{plain,right=0.75}{v1,va}
\fmf{plain}{va,v2}
\fmf{plain,left=0.75}{v1,va}
\fmf{plain}{va,v3}
\fmf{plain}{v2,v3}
\fmfv{decor.shape=circle,decor.filled=30,decor.size=9}{v1}
\fmfv{decor.shape=circle,decor.filled=30,decor.size=9}{v2}
\fmfv{decor.shape=circle,decor.filled=30,decor.size=9}{v3}
\end{fmfgraph*}}
~ +~
\parbox{20mm}{
\begin{fmfgraph*}(50,50)
\fmftop{v1}
\fmfbottom{v2,v3}
\fmf{plain,right=0.75}{v2,va}
\fmf{plain}{va,v1}
\fmf{plain,left=0.75}{v2,va}
\fmf{plain}{va,v3}
\fmf{plain}{v1,v3}
\fmfv{decor.shape=circle,decor.filled=30,decor.size=9}{v1}
\fmfv{decor.shape=circle,decor.filled=30,decor.size=9}{v2}
\fmfv{decor.shape=circle,decor.filled=30,decor.size=9}{v3}
\end{fmfgraph*}}
~+~
\parbox{20mm}{
\begin{fmfgraph*}(50,50)
\fmftop{v1}
\fmfbottom{v2,v3}
\fmf{plain,right=0.75}{v3,va}
\fmf{plain}{va,v2}
\fmf{plain,left=0.75}{v3,va}
\fmf{plain}{va,v1}
\fmf{plain}{v1,v2}
\fmfv{decor.shape=circle,decor.filled=30,decor.size=9}{v1}
\fmfv{decor.shape=circle,decor.filled=30,decor.size=9}{v2}
\fmfv{decor.shape=circle,decor.filled=30,decor.size=9}{v3}
\end{fmfgraph*}}
~\Biggr)
=2 ~
\parbox{20mm}{
\begin{fmfgraph*}(50,50)
\fmfleft{i}
\fmfright{o}
\fmf{plain,tension=5.4}{i,v1}
\fmf{plain,tension=5.4}{v2,o}
\fmf{plain, right,tension=1}{v1,va}
\fmf{plain, left,tension=1}{v1,va}
\fmf{dbl_plain, right,tension=1}{v2,va}
\fmf{plain, left,tension=1}{v2,va}
\end{fmfgraph*}}
\]
\vspace{0.4cm}
\[
\hspace{-4.5cm}
\int d^{2\o}x_1 \Biggl(\quad
\parbox{20mm}{
\begin{fmfgraph*}(50,50)
\fmftop{v1}
\fmfbottom{v2,v3}
\fmf{plain}{v1,va,v2}
\fmf{plain}{v1,v3}
\fmf{plain}{v2,v3}
\fmffreeze
\fmfv{d.sh=circle,l.d=0, d.f=empty,d.si=.3w,l=$1$}{va}
\fmfv{decor.shape=circle,decor.filled=30,decor.size=9}{v1}
\fmfv{decor.shape=circle,decor.filled=30,decor.size=9}{v2}
\fmfv{decor.shape=circle,decor.filled=30,decor.size=9}{v3}
\end{fmfgraph*}}
~+~
\parbox{20mm}{
\begin{fmfgraph*}(50,50)
\fmftop{v1}
\fmfbottom{v2,v3}
\fmf{plain}{v1,v2}
\fmf{plain}{v1,va,v3}
\fmf{plain}{v2,v3}
\fmffreeze
\fmfv{d.sh=circle,l.d=0, d.f=empty,d.si=.3w,l=$1$}{va}
\fmfv{decor.shape=circle,decor.filled=30,decor.size=9}{v1}
\fmfv{decor.shape=circle,decor.filled=30,decor.size=9}{v2}
\fmfv{decor.shape=circle,decor.filled=30,decor.size=9}{v3}
\end{fmfgraph*}}
~+~
\parbox{20mm}{
\begin{fmfgraph*}(50,50)
\fmftop{v1}
\fmfbottom{v2,v3}
\fmf{plain}{v1,v2}
\fmf{plain}{v1,v3}
\fmf{plain}{v2,va,v3}
\fmffreeze
\fmfv{d.sh=circle,l.d=0, d.f=empty,d.si=.3w,l=$1$}{va}
\fmfv{decor.shape=circle,decor.filled=30,decor.size=9}{v1}
\fmfv{decor.shape=circle,decor.filled=30,decor.size=9}{v2}
\fmfv{decor.shape=circle,decor.filled=30,decor.size=9}{v3}
\end{fmfgraph*}}
~\Biggr)
\]
\[
=-2~
\parbox{20mm}{
\begin{fmfgraph*}(60,60)
\fmfleft{i}
\fmfright{o}
\fmftop{t}
\fmf{plain,tension=1.4}{i,v1}
\fmf{plain,tension=1.4}{v2,o}
\fmf{dbl_plain, right,tension=1}{v1,v2}
\fmffreeze
\fmf{phantom, tension=.8}{t,v3}
\fmf{plain, right=0.4,tension=0.4}{v3,v1}
\fmf{plain, left=0.64,tension=0.4}{v3,v1}
\fmf{plain, right=0.4,tension=0.4}{v2,v3}
\end{fmfgraph*}}
~-4~
\parbox{20mm}{
\begin{fmfgraph*}(60,60)
\fmfleft{i}
\fmfright{o}
\fmftop{t}
\fmf{plain,tension=1.4}{i,v1}
\fmf{plain,tension=1.4}{v2,o}
\fmf{plain, right,tension=1}{v1,v2}
\fmffreeze
\fmf{phantom, tension=.8}{t,v3}
\fmf{plain, right=0.4,tension=0.4}{v3,v1}
\fmf{plain, left=0.64,tension=0.4}{v3,v1}
\fmf{dbl_plain, right=0.4,tension=0.4}{v2,v3}
\end{fmfgraph*}}
\]

\noindent We will also need to compute the normalization of the
two-point functions, given by the following diagrams
\[
\parbox{20mm}{
\hspace{-1.2cm}
\begin{fmfgraph*}(60,60)
\fmfleft{v1}
\fmfright{v3}
\fmf{plain,left=0.25, tension=1}{v1,v4,v3,v2,v1}
\fmffixed{(0,30)}{v2,v4}
\fmf{wiggly}{v2,v4}
\fmfv{decor.shape=circle,decor.filled=30,decor.size=9}{v1}
\fmfv{decor.shape=circle,decor.filled=30,decor.size=9}{v3}
\end{fmfgraph*}}
\hspace{-0.5cm} =4~
\parbox{20mm}{
\begin{fmfgraph*}(60,60)
\fmfleft{i}
\fmfright{o}
\fmftop{t}
\fmf{plain,tension=1.4}{i,v1}
\fmf{plain,tension=1.4}{v2,o}
\fmf{plain, right,tension=1}{v1,v2}
\fmffreeze
\fmf{phantom, tension=.8}{t,v3}
\fmf{plain, right=0.4,tension=0.4}{v3,v1}
\fmf{plain, left=0.64,tension=0.4}{v3,v1}
\fmf{plain, right=0.4,tension=0.4}{v2,v3}
\end{fmfgraph*}}
~ -2p^2~
\parbox{20mm}{
\begin{fmfgraph*}(50,50)
\fmfleft{i}
\fmfright{o}
\fmf{plain, tension=2}{i,v1}
\fmf{plain, tension=2}{v3,o}
\fmf{plain,left=0.4, tension=1}{v1,v4,v3,v2,v1}
\fmffixed{(0,26)}{v2,v4}
\fmf{plain}{v2,v4}
\end{fmfgraph*}}
-~
\parbox{20mm}{
\begin{fmfgraph*}(50,50)
\fmfleft{i}
\fmfright{o}
\fmf{plain,tension=5.4}{i,v1}
\fmf{plain,tension=5.4}{v2,o}
\fmf{plain, right,tension=1}{v1,va}
\fmf{plain, left,tension=1}{v1,va}
\fmf{plain, right,tension=1}{v2,va}
\fmf{plain, left,tension=1}{v2,va}
\end{fmfgraph*}}
\]
\vspace{-0.85cm}
\[
\hspace{-3.6cm}
\parbox{20mm}{
\begin{fmfgraph*}(60,60)
\fmfleft{v1}
\fmfright{v2}
\fmftop{t}
\fmf{plain, right=0.4,tension=.3}{v1,v2}
\fmf{phantom, tension=.8}{t,v3}
\fmf{plain, right=0.2,tension=0.54}{v3,v1}
\fmf{plain, right=0.2,tension=0.54}{v2,v3}
\fmfv{d.sh=circle,l.d=0, d.f=empty,d.si=.25w,l=$1$}{v3}
\fmfv{decor.shape=circle,decor.filled=30,decor.size=9}{v1}
\fmfv{decor.shape=circle,decor.filled=30,decor.size=9}{v2}
\end{fmfgraph*}}
\quad+~~
\parbox{20mm}{
\begin{fmfgraph*}(60,60)
\fmfleft{v1}
\fmfright{v2}
\fmfbottom{b}
\fmf{plain, left=0.4,tension=.3}{v1,v2}
\fmf{phantom, tension=.6}{b,v3}
\fmf{plain, left=0.2,tension=0.54}{v3,v1}
\fmf{plain, left=0.2,tension=0.54}{v2,v3}
\fmfv{d.sh=circle,l.d=0, d.f=empty,d.si=.25w,l=$1$}{v3}
\fmfv{decor.shape=circle,decor.filled=30,decor.size=9}{v1}
\fmfv{decor.shape=circle,decor.filled=30,decor.size=9}{v2}
\end{fmfgraph*}}
\quad~ =-4~
\parbox{20mm}{
\begin{fmfgraph*}(60,60)
\fmfleft{i}
\fmfright{o}
\fmftop{t}
\fmf{plain,tension=1.4}{i,v1}
\fmf{plain,tension=1.4}{v2,o}
\fmf{plain, right,tension=1}{v1,v2}
\fmffreeze
\fmf{phantom, tension=.8}{t,v3}
\fmf{plain, right=0.4,tension=0.4}{v3,v1}
\fmf{plain, left=0.64,tension=0.4}{v3,v1}
\fmf{plain, right=0.4,tension=0.4}{v2,v3}
\end{fmfgraph*}}
\]
\vspace{-0.85cm}
\[
\hspace{-7.4cm}
\parbox{20mm}{
\begin{fmfgraph*}(60,60)
\fmfleft{v1}
\fmfright{v2}
\fmf{plain, right=0.75,tension=1}{v1,va}
\fmf{plain, left=0.75,tension=1}{v1,va}
\fmf{plain, right=0.75,tension=1}{v2,va}
\fmf{plain, left=0.75,tension=1}{v2,va}
\fmfv{decor.shape=circle,decor.filled=30,decor.size=9}{v1}
\fmfv{decor.shape=circle,decor.filled=30,decor.size=9}{v2}
\end{fmfgraph*}}
\quad=~
\parbox{20mm}{
\begin{fmfgraph*}(50,50)
\fmfleft{i}
\fmfright{o}
\fmf{plain,tension=5.4}{i,v1}
\fmf{plain,tension=5.4}{v2,o}
\fmf{plain, right,tension=1}{v1,va}
\fmf{plain, left,tension=1}{v1,va}
\fmf{plain, right,tension=1}{v2,va}
\fmf{plain, left,tension=1}{v2,va}
\end{fmfgraph*}}
\]

\noindent Summing-up these diagrams we find the following results
\[
N\int d^{2\o}x_1 \bigl\la {\cal O}(x_1) {\cal O}(x_2) {\cal O}(x_3)
\bigr\ra_{{\cal O}(\l)} = \int \frac{d^{2\o}p}{(2\pi)^{2\o}} e^{ip\cdot x_{23}} \Bigl(
4~
\parbox{20mm}{
\begin{fmfgraph*}(60,60)
\fmfleft{i}
\fmfright{o}
\fmftop{t}
\fmf{plain,tension=1.4}{i,v1}
\fmf{plain,tension=1.4}{v2,o}
\fmf{plain, right,tension=1}{v1,v2}
\fmffreeze
\fmf{phantom, tension=.8}{t,v3}
\fmf{dbl_plain, right=0.4,tension=0.4}{v3,v1}
\fmf{plain, left=0.64,tension=0.4}{v3,v1}
\fmf{plain, right=0.4,tension=0.4}{v2,v3}
\end{fmfgraph*}}
\]
\vspace{-.75cm}
\[
+2~
\parbox{20mm}{
\begin{fmfgraph*}(50,50)
\fmfleft{i}
\fmfright{o}
\fmf{plain, tension=2}{i,v1}
\fmf{plain, tension=2}{v3,o}
\fmf{plain,left=0.4, tension=1}{v1,v4,v3,v2,v1}
\fmffixed{(0,26)}{v2,v4}
\fmf{plain}{v2,v4}
\end{fmfgraph*}}
-4p^2~
\parbox{20mm}{
\begin{fmfgraph*}(50,50)
\fmfleft{i}
\fmfright{o}
\fmf{plain, tension=2}{i,v1}
\fmf{plain, tension=2}{v3,o}
\fmf{plain,left=0.4, tension=1}{v1,v4,v3,v2,v1}
\fmffixed{(0,26)}{v2,v4}
\fmf{plain}{v2,v4}
\fmffreeze
\fmf{dbl_plain,left=0.4, tension=1}{v1,v4}
\end{fmfgraph*}}
\Bigr),
\]
\vspace{-0.4cm}
\[
\hspace{-1.75cm}
\bigl\la {\cal O}_i(x) {\cal O}_i(0) \bigr\ra_{{\cal O}(\l)} = 
\int \frac{d^{2\o}p}{(2\pi)^{2\o}} e^{ip\cdot x} 
\Bigl( -2p^2~
\parbox{20mm}{
\begin{fmfgraph*}(50,50)
\fmfleft{i}
\fmfright{o}
\fmf{plain, tension=2}{i,v1}
\fmf{plain, tension=2}{v3,o}
\fmf{plain,left=0.4, tension=1}{v1,v4,v3,v2,v1}
\fmffixed{(0,26)}{v2,v4}
\fmf{plain}{v2,v4}
\end{fmfgraph*}}
\Bigr).
\]
\noindent These expressions evaluate to
\bsp
&N\int d^{2\o}x_1 \bigl\la {\cal O}_1(x_1) {\cal O}_2(x_2) {\cal O}_3(x_3)
\bigr\ra_{{\cal O}(\l)} =
\frac{1}{x_{23}^2}
\frac{3\zeta(3)}{2^8\pi^6} + {\cal O}(\e),\\
&\bigl\la {\cal O}_i(x) {\cal O}_i(0) \bigr\ra_{{\cal O}(\l)} = -\e
\frac{3\zeta(3)}{64\pi^6}\frac{1}{(x^2)^2} +{\cal O}(\e^2).
\end{split}
\ee
Therefore we have from (\ref{C}) and (\ref{g}) that
\be
\hat C_{123}\bigl|_{{\cal O}(\l)} =
g_i\bigl|_{{\cal O}(\l)} = -3\e\frac{\zeta(3)}{4\pi^2},\qquad C_{123}\bigl|_{{\cal O}(\l)}=
3\e\frac{\zeta(3)}{8\pi^2} = {\cal O}(\e),
\ee
and so both $g_i$ and $\hat C_{123}$ are zero at the one-loop order
and hence trivially so is $C_{123}\bigl|_{{\cal O}(\l)}=0$ on the
physical dimension as expected.

\section{Main calculation and results}
\label{sec:main}

In this section we summarize the results of the ABJ(M)
calculation. The method used is that employed in
\cite{Young:2013hda}. Namely Feynman rules spelled-out in
\cite{Minahan:2009wg} are processed into master integrals using FIRE
\cite{Smirnov:2008iw}. Note that for ABJM we cannot use dimensional
reduction as there is no higher dimensional supersymmetric theory to
reduce from. The scheme used here is that employed successfully in
\cite{Minahan:2009wg}: i.e. to reduce all numerators to scalar
products in $d=3$ before integrating using $d=3-2\e$.  In tables
\ref{tab:3} and \ref{tab:2} we list all non-zero Feynman diagrams. The
results for these diagrams and further details of the calculation are
found in appendix \ref{app:det}.
\begin{table}
\begin{tabular}{cccccc}
\parbox{20mm}{
\vspace{0.5cm}
\begin{fmfgraph*}(50,50)
\fmftop{v1}
\fmfbottom{v2,v3}
\fmf{plain}{v1,va,v2}
\fmf{plain}{v1,vb,v3}
\fmf{plain}{v2,v3}
\fmffreeze
\fmf{wiggly,left=0.3}{va,vb}
\fmf{wiggly,right=0.3}{va,vb}
\fmfv{decor.shape=circle,decor.filled=30,decor.size=9}{v1}
\fmfv{decor.shape=circle,decor.filled=30,decor.size=9}{v2}
\fmfv{decor.shape=circle,decor.filled=30,decor.size=9}{v3}
\end{fmfgraph*}}&
\parbox{20mm}{
\vspace{0.5cm}
\begin{fmfgraph*}(50,50)
\fmftop{v1}
\fmfbottom{v2,v3}
\fmf{plain}{v1,va,v2}
\fmf{plain}{v1,vb,vc,v3}
\fmf{plain}{v2,v3}
\fmffreeze
\fmf{wiggly}{va,vb}
\fmf{wiggly}{va,vc}
\fmfv{decor.shape=circle,decor.filled=30,decor.size=9}{v1}
\fmfv{decor.shape=circle,decor.filled=30,decor.size=9}{v2}
\fmfv{decor.shape=circle,decor.filled=30,decor.size=9}{v3}
\end{fmfgraph*}}&
\parbox{20mm}{
\vspace{0.5cm}
\begin{fmfgraph*}(50,50)
\fmftop{v1}
\fmfbottom{v2,v3}
\fmf{plain}{v1,va,vb,vc,v2}
\fmf{plain}{v1,vd,v3}
\fmf{plain}{v2,v3}
\fmffreeze
\fmf{wiggly,right=0.7}{va,vc}
\fmf{wiggly}{vb,vd}
\fmfv{decor.shape=circle,decor.filled=30,decor.size=9}{v1}
\fmfv{decor.shape=circle,decor.filled=30,decor.size=9}{v2}
\fmfv{decor.shape=circle,decor.filled=30,decor.size=9}{v3}
\end{fmfgraph*}}&
\parbox{20mm}{
\vspace{0.5cm}
\begin{fmfgraph*}(50,50)
\fmftop{v1}
\fmfbottom{v2,v3}
\fmf{plain}{v1,va,vb,v2}
\fmf{plain}{v1,vc,v3}
\fmf{plain}{v2,v3}
\fmffreeze
\fmf{wiggly,left=0.5,tension=0.5}{va,vd,vb}
\fmf{wiggly,tension=0.5}{vd,vc}
\fmfv{decor.shape=circle,decor.filled=30,decor.size=9}{v1}
\fmfv{decor.shape=circle,decor.filled=30,decor.size=9}{v2}
\fmfv{decor.shape=circle,decor.filled=30,decor.size=9}{v3}
\end{fmfgraph*}}&
\parbox{20mm}{
\vspace{0.5cm}
\begin{fmfgraph*}(50,50)
\fmftop{v1}
\fmfbottom{v2,v3}
\fmf{plain}{v1,va,v2}
\fmf{plain}{v1,vb,v3}
\fmf{plain}{v2,v3}
\fmffreeze
\fmf{wiggly}{va,vc,vb}
\fmfv{d.sh=circle,l.d=0, d.f=empty,d.si=.25w,l=$1$}{vc}
\fmfv{decor.shape=circle,decor.filled=30,decor.size=9}{v1}
\fmfv{decor.shape=circle,decor.filled=30,decor.size=9}{v2}
\fmfv{decor.shape=circle,decor.filled=30,decor.size=9}{v3}
\end{fmfgraph*}}&
\parbox{20mm}{
\vspace{0.5cm}
\begin{fmfgraph*}(50,50)
\fmftop{v1}
\fmfbottom{v2,v3}
\fmf{plain,right=0.75}{v1,va}
\fmf{dashes,right=0.75}{va,vb}
\fmf{plain}{vb,v2}
\fmf{plain,left=0.75}{v1,va}
\fmf{dashes,left=0.75}{va,vb}
\fmf{plain}{vb,v3}
\fmf{plain}{v2,v3}
\fmfv{decor.shape=circle,decor.filled=30,decor.size=9}{v1}
\fmfv{decor.shape=circle,decor.filled=30,decor.size=9}{v2}
\fmfv{decor.shape=circle,decor.filled=30,decor.size=9}{v3}
\end{fmfgraph*}}\\
\parbox{20mm}{
\vspace{0.5cm}
\begin{fmfgraph*}(50,50)
\fmftop{v1}
\fmfbottom{v2,v3}
\fmf{plain}{v1,va,v2}
\fmf{plain}{v1,vb,v3}
\fmf{plain}{v2,v3}
\fmffreeze
\fmf{dashes,left=0.3}{va,vb}
\fmf{dashes,right=0.3}{va,vb}
\fmfv{decor.shape=circle,decor.filled=30,decor.size=9}{v1}
\fmfv{decor.shape=circle,decor.filled=30,decor.size=9}{v2}
\fmfv{decor.shape=circle,decor.filled=30,decor.size=9}{v3}
\end{fmfgraph*}}&
\parbox{20mm}{
\vspace{0.5cm}
\begin{fmfgraph*}(50,50)
\fmftop{v1}
\fmfbottom{v2,v3}
\fmf{plain}{v1,va,vd,v2}
\fmf{plain}{v1,vb,vc,v3}
\fmf{plain}{v2,v3}
\fmffreeze
\fmf{wiggly}{va,vc}
\fmf{wiggly}{vb,vd}
\fmfv{decor.shape=circle,decor.filled=30,decor.size=9}{v1}
\fmfv{decor.shape=circle,decor.filled=30,decor.size=9}{v2}
\fmfv{decor.shape=circle,decor.filled=30,decor.size=9}{v3}
\end{fmfgraph*}}&
\parbox{20mm}{
\vspace{0.5cm}
\begin{fmfgraph*}(50,50)
\fmftop{v1}
\fmfbottom{v2,v3}
\fmf{plain}{v1,va,v2}
\fmf{plain}{v1,vb,v3}
\fmf{plain}{v2,vd,vc,v3}
\fmffreeze
\fmf{wiggly}{va,vc}
\fmf{wiggly}{vb,vd}
\fmfv{decor.shape=circle,decor.filled=30,decor.size=9}{v1}
\fmfv{decor.shape=circle,decor.filled=30,decor.size=9}{v2}
\fmfv{decor.shape=circle,decor.filled=30,decor.size=9}{v3}
\end{fmfgraph*}}&
\parbox{20mm}{
\vspace{0.5cm}
\begin{fmfgraph*}(50,50)
\fmftop{v1}
\fmfbottom{v2,v3}
\fmf{plain}{v1,va,v2}
\fmf{plain}{v1,vb,v3}
\fmf{plain}{v2,vc,v3}
\fmffreeze
\fmf{wiggly}{va,vb}
\fmf{wiggly}{va,vc}
\fmfv{decor.shape=circle,decor.filled=30,decor.size=9}{v1}
\fmfv{decor.shape=circle,decor.filled=30,decor.size=9}{v2}
\fmfv{decor.shape=circle,decor.filled=30,decor.size=9}{v3}
\end{fmfgraph*}}&
\parbox{20mm}{
\vspace{0.5cm}
\begin{fmfgraph*}(50,50)
\fmftop{v1}
\fmfbottom{v2,v3}
\fmf{plain}{v1,va,v2}
\fmf{plain}{v1,vc,v3}
\fmf{plain}{v2,vb,v3}
\fmffreeze
\fmf{wiggly,tension=0.5}{va,vd}
\fmf{wiggly,tension=0.5}{vb,vd}
\fmf{wiggly,tension=0.5}{vc,vd}
\fmfv{decor.shape=circle,decor.filled=30,decor.size=9}{v1}
\fmfv{decor.shape=circle,decor.filled=30,decor.size=9}{v2}
\fmfv{decor.shape=circle,decor.filled=30,decor.size=9}{v3}
\end{fmfgraph*}}&
\parbox{20mm}{
\vspace{0.5cm}
\begin{fmfgraph*}(50,50)
\fmftop{v1}
\fmfbottom{v2,v3}
\fmf{plain}{v1,va,v2}
\fmf{plain}{v1,v3}
\fmf{plain}{v2,v3}
\fmfv{d.sh=circle,l.d=0, d.f=empty,d.si=.25w,l=$2$}{va}
\fmfv{decor.shape=circle,decor.filled=30,decor.size=9}{v1}
\fmfv{decor.shape=circle,decor.filled=30,decor.size=9}{v2}
\fmfv{decor.shape=circle,decor.filled=30,decor.size=9}{v3}
\end{fmfgraph*}}
\end{tabular}
\caption{The Feynman diagrams which contribute to the three-point
  function in ABJ(M). Note that all unique diagrams obtained from
  these via rotations (by $2\pi/3$) and reflections about the
  perpendicular bisectors of the main triangle must also be considered.}
\label{tab:3}
\end{table}
\begin{table}
\begin{tabular}{ccccc}
\parbox{20mm}{
\begin{fmfgraph*}(50,50)
\fmfleft{v1}
\fmfright{v3}
\fmf{plain,left=0.3, tension=1}{v1,v4,v3,v2,v1}
\fmffixed{(0,33)}{v2,v4}
\fmf{wiggly,left=.3}{v2,v4,v2}
\fmfv{decor.shape=circle,decor.filled=30,decor.size=9}{v1}
\fmfv{decor.shape=circle,decor.filled=30,decor.size=9}{v3}
\end{fmfgraph*}}&
\parbox{20mm}{
\begin{fmfgraph*}(50,50)
\fmfleft{v1}
\fmftop{t}
\fmfright{v2}
\fmfbottom{b1,b2}
\fmf{phantom,tension=2.95}{t,va}
\fmf{phantom,tension=2.15}{b1,vb1}
\fmf{phantom,tension=2.15}{b2,vb2}
\fmf{plain,left=0.3}{v1,va}
\fmf{plain,left=0.3}{va,v2}
\fmf{plain,right=0.25}{v1,vb1}
\fmf{plain,right=0.25}{vb1,vb2}
\fmf{plain,right=0.25}{vb2,v2}
\fmffreeze
\fmf{wiggly}{va,vb1}
\fmf{wiggly}{va,vb2}
\fmfv{decor.shape=circle,decor.filled=30,decor.size=9}{v1}
\fmfv{decor.shape=circle,decor.filled=30,decor.size=9}{v2}
\end{fmfgraph*}}&
\parbox{20mm}{
\begin{fmfgraph*}(50,50)
\fmfleft{v1}
\fmftop{t1,t2,t3}
\fmfright{v2}
\fmfbottom{b}
\fmf{phantom,tension=1.0}{t1,va1}
\fmf{phantom,tension=1.1}{t2,va2}
\fmf{phantom,tension=1.0}{t3,va3}
\fmf{phantom,tension=2.95}{b,vb}
\fmf{plain,left=0.17}{v1,va1}
\fmf{plain,left=0.15}{va1,va2}
\fmf{plain,left=0.15}{va2,va3}
\fmf{plain,left=0.17}{va3,v2}
\fmf{plain,right=0.3}{v1,vb}
\fmf{plain,right=0.3}{vb,v2}
\fmffreeze
\fmf{wiggly,left=0.75}{va1,va3}
\fmf{wiggly}{va2,vb}
\fmfv{decor.shape=circle,decor.filled=30,decor.size=9}{v1}
\fmfv{decor.shape=circle,decor.filled=30,decor.size=9}{v2}
\end{fmfgraph*}}&
\parbox{20mm}{
\begin{fmfgraph*}(50,50)
\fmfleft{v1}
\fmftop{t1,t2,t3}
\fmfright{v2}
\fmfbottom{b}
\fmf{phantom,tension=1.0}{t1,va1}
\fmf{phantom,tension=1.1}{t2,va2}
\fmf{phantom,tension=1.0}{t3,va3}
\fmf{phantom,tension=2.95}{b,vb}
\fmf{plain,left=0.17}{v1,va1}
\fmf{plain,left=0.15}{va1,va2}
\fmf{plain,left=0.15}{va2,va3}
\fmf{plain,left=0.17}{va3,v2}
\fmf{plain,right=0.3}{v1,vb}
\fmf{plain,right=0.3}{vb,v2}
\fmffreeze
\fmf{wiggly}{va1,vcen}
\fmf{wiggly}{va3,vcen}
\fmf{wiggly}{vb,vcen}
\fmfv{decor.shape=circle,decor.filled=30,decor.size=9}{v1}
\fmfv{decor.shape=circle,decor.filled=30,decor.size=9}{v2}
\end{fmfgraph*}}&
\parbox{20mm}{
\begin{fmfgraph*}(50,50)
\fmfleft{v1}
\fmfright{v3}
\fmf{plain,left=0.3, tension=1}{v1,v4,v3,v2,v1}
\fmffixed{(0,33)}{v2,v4}
\fmf{wiggly}{v2,vc,v4}
\fmfv{d.sh=circle,l.d=0, d.f=empty,d.si=.25w,l=$1$}{vc}
\fmfv{decor.shape=circle,decor.filled=30,decor.size=9}{v1}
\fmfv{decor.shape=circle,decor.filled=30,decor.size=9}{v3}
\end{fmfgraph*}}\\
\parbox{20mm}{
\begin{fmfgraph*}(50,50)
\fmfleft{v1}
\fmfright{v3}
\fmf{plain,left=1., tension=1}{v1,va}
\fmf{dashes,left=1., tension=1}{va,vb}
\fmf{plain,left=1., tension=1}{vb,v3}
\fmf{plain,right=1., tension=1}{v1,va}
\fmf{dashes,right=1., tension=1}{va,vb}
\fmf{plain,right=1., tension=1}{vb,v3}
\fmfv{decor.shape=circle,decor.filled=30,decor.size=9}{v1}
\fmfv{decor.shape=circle,decor.filled=30,decor.size=9}{v3}
\end{fmfgraph*}}&
\parbox{20mm}{
\begin{fmfgraph*}(50,50)
\fmfleft{v1}
\fmfright{v3}
\fmf{plain,left=0.3, tension=1}{v1,v4,v3,v2,v1}
\fmffixed{(0,33)}{v2,v4}
\fmf{dashes,left=.3}{v2,v4,v2}
\fmfv{decor.shape=circle,decor.filled=30,decor.size=9}{v1}
\fmfv{decor.shape=circle,decor.filled=30,decor.size=9}{v3}
\end{fmfgraph*}}&
\parbox{20mm}{
\begin{fmfgraph*}(50,50)
\fmfleft{v1}
\fmftop{t1,t2}
\fmfright{v2}
\fmfbottom{b1,b2}
\fmf{phantom,tension=1.75}{t1,va1}
\fmf{phantom,tension=1.75}{t2,va2}
\fmf{phantom,tension=1.75}{b1,vb1}
\fmf{phantom,tension=1.75}{b2,vb2}
\fmf{plain,left=0.25}{v1,va1}
\fmf{plain,left=0.25}{va1,va2}
\fmf{plain,left=0.25}{va2,v2}
\fmf{plain,right=0.25}{v1,vb1}
\fmf{plain,right=0.25}{vb1,vb2}
\fmf{plain,right=0.25}{vb2,v2}
\fmffreeze
\fmf{wiggly}{va1,vb2}
\fmf{wiggly}{va2,vb1}
\fmfv{decor.shape=circle,decor.filled=30,decor.size=9}{v1}
\fmfv{decor.shape=circle,decor.filled=30,decor.size=9}{v2}
\end{fmfgraph*}}&
\parbox{20mm}{
\begin{fmfgraph*}(50,50)
\fmfleft{v1}
\fmfright{v3}
\fmf{plain,left=0.3, tension=1}{v1,v4,v3,v2,v1}
\fmffixed{(0,30)}{v2,v4}
\fmfv{d.sh=circle,l.d=0, d.f=empty,d.si=.25w,l=$2$}{v4}
\fmfv{decor.shape=circle,decor.filled=30,decor.size=9}{v1}
\fmfv{decor.shape=circle,decor.filled=30,decor.size=9}{v3}
\end{fmfgraph*}}&
\end{tabular}
\caption{The Feynman diagrams which contribute to the two-point
  function in ABJ(M). Unique diagrams obtained through reflection
  about the horizontal axis must also be considered.}
\label{tab:2}
\end{table}
The results are as follows (below we quote the two-loop corrections to
$C_{123}$, $\hat C_{123}$ and $g$; they are equal to unity at tree-level)
\bsp
&\sqrt{MN}\int d^{2\o}x_1 \bigl\la {\cal O}_1(x_1) {\cal O}_2(x_2) {\cal O}_3(x_3)
\bigr\ra =-\frac{\hat C_{123}}{32\pi^2}=
(\l+\hat\l)^2\frac{1}{2^{10}}\left(\frac{5}{3}+\frac{8}{\pi^2}\right)\\
&\qquad\qquad\qquad\qquad\qquad\qquad
+(\l-\hat\l)^2\frac{1}{2^{10}}\left(\frac{5}{3}-\frac{24}{\pi^2}\right)
+ {\cal O}(\e),\\
&\bigl\la {\cal O}_i(x) {\cal O}_i(0) \bigr\ra =\frac{g}{16\pi^2x^2}=
-\frac{\l^2+\hat\l^2}{96}\frac{1}{x^2} 
+{\cal O}(\e).
\end{split}
\ee
We notice here that the two-point function is finite, as expected, and
that factors of Euler's constant $\g$ and $\log\pi$ cancel out. This we
interpret as confirmation of regularization scheme independence. We
thus find
\bsp\label{bres}
C_{123} &= -32\pi^2\sqrt{MN} \int d^{2\o}x_1 
\bigl\la {\cal O}_1(x_1) {\cal O}_2(x_2) {\cal O}_3(x_3)
\bigr\ra - \frac{3g}{2} \\ 
&=  \boxed{
(\l+\hat\l)^2 \left(\frac{7\pi^2}{96}-\frac{1}{4}\right) 
+(\l-\hat\l)^2 \left(\frac{7\pi^2}{96}+\frac{3}{4}\right) .}
\end{split}
\ee

\section{Discussion}
\label{sec:disc}

We have computed the structure constant for a CPO three-point function
in ABJ(M) theory at leading order in perturbation theory
(\ref{bres}). Concentrating on the ABJM case, we now have results for
this quantity both at weak and at strong coupling in the planar limit;
indeed using (\ref{sc}) we have
\be
C_{123} (\l \ll 1) = 1 -
\l^2\left(1-\frac{7\pi^2}{24}\right),\qquad
C_{123} (\l \gg 1) = \left(\frac{\l}{2} \right)^{1/4} \frac{\sqrt{3\pi}}{2}.
\ee
It would be very interesting to find a way to compute this
interpolating function for all values of the 't Hooft
coupling. Although the machinery of integrability has been applied to
computing three-point functions in ${\cal N}=4$ SYM
\cite{Escobedo:2010xs,Escobedo:2011xw,Gromov:2011jh,Gromov:2012uv},
here we have a rather different situation, in that three spin-chain
vacuum states, i.e. states which are annihilated by the dilatation
operator, nevertheless have a non-trivial structure function dependent
upon the 't Hooft coupling. Indeed, because of the fact that at
tree-level the tensors defining the CPO's contract in just one way,
whereas at strong coupling they contract in all ways, it is clear that
there are a host of interpolating functions, and each one likely
begins at a different order in the 't Hooft coupling in the
perturbative expansion. Understanding the connections between these
functions remains a very interesting direction of future research.

We know that the M-theory dual of ABJ theory involves the appearance
of a three-form in an $S^3/\mathbb{Z}_k\subset S^7/\mathbb{Z}_k$
\cite{Aharony:2008gk}. It would be interesting to compute the
fluctuation spectrum around this background and repeat the three-point
function calculation in order to have a version of (\ref{sc}) with
$N\neq M$.

\section*{Acknowledgements}

It is a pleasure to thank Andreas Brandhuber, Jan Plefka, Sanjaye
Ramgoolam, Rodolfo Russo, Gordon Semenoff, Gabriele Travaglini, Brian
Wecht, and Konstantin Zarembo for discussions.

\appendix

\section{Calculational details}
\label{app:det}

We use the machinery developed in \cite{Minahan:2009wg} and employed
in a very similar setting, the calculation of the two-loop form factor
of CPO's, in \cite{Young:2013hda}. These references contain ample
detail and we choose not to reprint the details of the action, Feynman
rules, etc. here but rather refer the interested reader to these
papers.

\subsection{Master integrals}

The three-loop master integrals required to complete the calculation
are as follows
\vspace{-0.5cm}
\[P_1 = 
\parbox{20mm}{
\begin{fmfgraph*}(70,70)
\fmfleft{i}
\fmfright{o}
\fmf{plain,tension=4}{i,v1}
\fmf{plain,tension=4}{v3,o}
\fmf{plain,left=1}{v1,va,v3,va,v1}
\fmffreeze
\fmf{plain}{v1,va}
\end{fmfgraph*}}
\quad
=G^2(1,1)\,G(1,2-\o),\quad
P_2=
\parbox{20mm}{
\begin{fmfgraph*}(50,50)
\fmfleft{i}
\fmfright{o}
\fmf{plain,tension=4}{i,v1}
\fmf{plain,tension=4}{v3,o}
\fmf{plain,left=1}{v1,va,v3,va,v1}
\fmffreeze
\fmf{plain,left=1.}{v1,v3}
\end{fmfgraph*}}
=G^2(1,1)\,G(1,4-2\o),\]
\vspace{-1cm}
\[P_3=
\parbox{20mm}{
\begin{fmfgraph*}(60,60)
\fmfleft{i}
\fmfright{o}
\fmftop{t}
\fmfbottom{b}
\fmf{phantom,tension=1.7}{t,va}
\fmf{phantom,tension=1.7}{b,vb}
\fmf{plain,tension=4}{i,v1}
\fmf{plain,tension=4}{v3,o}
\fmf{plain,left=0.3}{v1,va,v3,vb,v1}
\fmffreeze
\fmf{plain,right=0.5}{va,vb,va}
\end{fmfgraph*}}
\,=G(1,1)\,F_{2-\o},~
P_4=
\parbox{20mm}{
\begin{fmfgraph*}(60,60)
\fmfleft{i}
\fmfright{o}
\fmf{plain,tension=8}{i,v1}
\fmf{plain,tension=8}{v3,o}
\fmf{plain,left=1}{v1,v3,v1}
\fmf{plain,left=0.5}{v1,v3,v1}
\end{fmfgraph*}}
\,=G(1,1)\,G(1,2-\o)\,G(1,3-2\o),\]
\vspace{-0.5cm}
\[
\hspace{-0.5cm}
P_5=
\parbox{20mm}{
\begin{fmfgraph*}(50,50)
\fmfleft{i}
\fmftop{t1,t2}
\fmfright{o}
\fmfbottom{b1,b2}
\fmf{plain,tension=4}{i,v1}
\fmf{plain,tension=4}{v2,o}
\fmf{phantom,tension=1.25}{t1,va1}
\fmf{phantom,tension=1.25}{t2,va2}
\fmf{phantom,tension=1.25}{b1,vb1}
\fmf{phantom,tension=1.25}{b2,vb2}
\fmf{plain,left=0.25}{v1,va1}
\fmf{plain,left=0.25}{va1,va2}
\fmf{plain,left=0.25}{va2,v2}
\fmf{plain,right=0.25}{v1,vb1}
\fmf{plain,right=0.25}{vb1,vb2}
\fmf{plain,right=0.25}{vb2,v2}
\fmffreeze
\fmf{plain}{va1,vb2}
\fmf{plain}{va2,vb1}
\end{fmfgraph*}}=\frac{1}{96}-\frac{13}{64\pi^2}+{\cal O}(\e),\quad
P_7=
\parbox{20mm}{
\begin{fmfgraph*}(50,50)
\fmfleft{i}
\fmfright{o}
\fmf{plain,tension=4}{i,v1}
\fmf{plain,tension=4}{v3,o}
\fmf{plain,tension=3}{v1,vc}
\fmf{plain,tension=3}{vd,v3}
\fmf{plain,left=1}{vc,va,vd,va,vc}
\fmffreeze
\fmf{plain,left=.9}{v1,v3}
\end{fmfgraph*}}\,=G^2(1,1)\,G(1,6-2\o),\]
\vspace{-0.5cm}
\[
\hspace{-5.95cm}
P_6=
\parbox{20mm}{
\begin{fmfgraph*}(60,60)
\fmfleft{i}
\fmfright{o}
\fmf{plain,tension=4}{i,v1}
\fmf{plain,tension=4}{v3,o}
\fmf{plain,left=1}{v1,v3}
\fmf{plain,tension=2}{v1,va}
\fmf{plain}{va,vb}
\fmf{plain,tension=2}{vb,v3}
\fmffreeze
\fmf{plain,left=1.}{va,vb,va}
\end{fmfgraph*}}\,=
G(1,1)\,G(1,2-\o)\,G(1,5-2\o),\]
where \cite{Kotikov:2003tc,Grozin:2003ak}
\bsp
&F_{\l} = \frac{2}{(4\pi)^{2\o}}\,\G(\o-1)\,\G(\o-\l-1)\,\G(\l-2\o+3)\Biggl(
-\frac{\pi\cot\left(\pi(2\o-\l)\right)}{\G(2\o-2)}\\
&+\frac{\G(\o-1)\,{}_3F_2\left(1,2+\l-\o,2\o-2;\l+1,\l-\o+3;1\right)}
{(\o-\l-2)\,\G(1+\l)\,\G(3\o-\l-4)}\,
\Biggr),
\end{split}
\ee
and where
\be\label{Gdef}
G(\a,\b) = \frac{1}{(4\pi)^\o}
\frac{\G(\a+\b-\o)\,\G(\o-\a)\,\G(\o-\b)}
{\G(\a)\,\G(\b)\,\G(2\o-\a-\b)}.
\ee
The non-planar integral $P_5$ must be evaluated using the Gegenbauer
polynomial technique \cite{Chetyrkin:1980pr},
c.f. \cite{Bekavac:2005xs}. The Fourier transform is given by 
\begin{equation}
\int \frac{d^{2\o}p}{(2\pi)^{2\o} } \frac{ e^{ip\cdot x} }{ [p^2]^s }
=\frac{\Gamma(\o-s)}{4^s\pi^\o\Gamma(s) }\frac{1}{[x^2]^{\o-s}}.
\end{equation} 

\subsection{Integral reduction for ABJ(M) diagrams}

We give here the results for the integrated three-point and two-point
diagrams in terms of basis integrals. Note that below a single
triangle diagram represents all unique diagrams obtained through
reflection and rotation, similarly a given two-point diagram
represents all unique diagrams gotten through reflections. A factor of
$(4\pi/k)^2$ is suppressed while the colour factors associated with the
various diagrams are as follows
\bsp
&I_7,T_7 \to -2MN,\quad
I_1,I_2,I_{10},T_1,T_2 \to (N-M)^2 - 2MN, \\
&I_3,I_5,I_8,I_9,T_3,T_5,T_8 \to 2MN, \quad
I_4,I_{11},T_4 \to N^2+M^2,\\
&I_6,T_6 \to (N-M)^2.
\end{split}
\ee
The two-loop self energy of the scalar field is given by
\cite{Minahan:2009wg}
\bsp
Z_\text{scalar}& = -\frac{1}{(4\pi)^2} \Biggl[
MN\left(\frac{3}{4(3-2\o)}+\frac{1}{4}\left(
-\frac{3\pi^2}{2} +25 -3\g +3\log(4\pi) \right)\right)\\
&+\frac{(M-N)^2}{4}\left(\frac{1}{2(3-2\o)}
-\frac{\pi^2}{4}+\frac{1}{2}\left(
3-\g+\log(4\pi)\right)\right)\Biggr]+{\cal O}(\e).
\end{split}
\ee

\subsubsection{Three-point diagrams}

\vspace{1cm}
\[
\hspace{-1.5cm}
I_1=-\frac{1}{4}I_7=
\int d^{2\o}x_1 \parbox{20mm}{
\begin{fmfgraph*}(50,50)
\fmftop{v1}
\fmfbottom{v2,v3}
\fmf{plain}{v1,va,v2}
\fmf{plain}{v1,vb,v3}
\fmf{plain}{v2,v3}
\fmffreeze
\fmf{wiggly,left=0.3}{va,vb}
\fmf{wiggly,right=0.3}{va,vb}
\fmfv{decor.shape=circle,decor.filled=30,decor.size=9}{v1}
\fmfv{decor.shape=circle,decor.filled=30,decor.size=9}{v2}
\fmfv{decor.shape=circle,decor.filled=30,decor.size=9}{v3}
\end{fmfgraph*}} 
=-\frac{1}{4}\int d^{2\o}x_1
\parbox{20mm}{
\begin{fmfgraph*}(50,50)
\fmftop{v1}
\fmfbottom{v2,v3}
\fmf{plain}{v1,va,v2}
\fmf{plain}{v1,vb,v3}
\fmf{plain}{v2,v3}
\fmffreeze
\fmf{dashes,left=0.3}{va,vb}
\fmf{dashes,right=0.3}{va,vb}
\fmfv{decor.shape=circle,decor.filled=30,decor.size=9}{v1}
\fmfv{decor.shape=circle,decor.filled=30,decor.size=9}{v2}
\fmfv{decor.shape=circle,decor.filled=30,decor.size=9}{v3}
\end{fmfgraph*}} 
= -\frac{(\omega -1)^2}{2 (\omega -2)} P_6\]\[+
\left(\omega -\frac{3}{2}\right)P_3 + \left(-12 \omega -\frac{3}{\omega -2}+11\right)P_4 
\]

\[
\hspace{-1.2cm}
I_2=\int d^{2\o}x_1 
\parbox{20mm}{
\begin{fmfgraph*}(50,50)
\fmftop{v1}
\fmfbottom{v2,v3}
\fmf{plain}{v1,va,v2}
\fmf{plain}{v1,vb,vc,v3}
\fmf{plain}{v2,v3}
\fmffreeze
\fmf{wiggly}{va,vb}
\fmf{wiggly}{va,vc}
\fmfv{decor.shape=circle,decor.filled=30,decor.size=9}{v1}
\fmfv{decor.shape=circle,decor.filled=30,decor.size=9}{v2}
\fmfv{decor.shape=circle,decor.filled=30,decor.size=9}{v3}
\end{fmfgraph*}}
=P_1 \left(-5 \omega +\frac{1}{\omega -2}+\frac{17}{2}\right)+P_2
\left(2 \omega +\frac{1}{4-2 \omega }-4\right)\]\[
+P_3 \left(\frac{19}{4}-3 \omega \right)+P_4 \left(24 \omega -\frac{5}{\omega -2}-\frac{3}{2 (\omega -2)^2}+\frac{1}{2 \omega -3}-32\right)
\]

\[
I_3=2I_4=\int d^{2\o}x_1 
\parbox{20mm}{
\begin{fmfgraph*}(50,50)
\fmftop{v1}
\fmfbottom{v2,v3}
\fmf{plain}{v1,va,vb,vc,v2}
\fmf{plain}{v1,vd,v3}
\fmf{plain}{v2,v3}
\fmffreeze
\fmf{wiggly,right=0.7}{va,vc}
\fmf{wiggly}{vb,vd}
\fmfv{decor.shape=circle,decor.filled=30,decor.size=9}{v1}
\fmfv{decor.shape=circle,decor.filled=30,decor.size=9}{v2}
\fmfv{decor.shape=circle,decor.filled=30,decor.size=9}{v3}
\end{fmfgraph*}}=2\int d^{2\o}x_1 
\parbox{20mm}{
\begin{fmfgraph*}(50,50)
\fmftop{v1}
\fmfbottom{v2,v3}
\fmf{plain}{v1,va,vb,v2}
\fmf{plain}{v1,vc,v3}
\fmf{plain}{v2,v3}
\fmffreeze
\fmf{wiggly,left=0.5,tension=0.5}{va,vd,vb}
\fmf{wiggly,tension=0.5}{vd,vc}
\fmfv{decor.shape=circle,decor.filled=30,decor.size=9}{v1}
\fmfv{decor.shape=circle,decor.filled=30,decor.size=9}{v2}
\fmfv{decor.shape=circle,decor.filled=30,decor.size=9}{v3}
\end{fmfgraph*}}
=-\frac{2 P_1 (5 \omega -8) (2 (\omega -4) \omega
  +7)}{(\omega -2)^2}\]\[
-P_2 \left(-24 \omega +\frac{6}{\omega -2}+\frac{2}{(\omega
  -2)^2}+40\right)
-P_3 \left(8 \omega +\frac{1}{\omega -2}-10\right)\]
\[-2 P_4 \left(-12
\omega -\frac{6}{\omega -2}+\frac{1}{3-2 \omega }+5\right)
-P_6 \left(-4 \omega -\frac{2}{\omega -2}+2\right)-2 P_7 (\omega -1)
\]

\[
I_5=\int d^{2\o}x_1 
\parbox{20mm}{
\begin{fmfgraph*}(50,50)
\fmftop{v1}
\fmfbottom{v2,v3}
\fmf{plain}{v1,va,v2}
\fmf{plain}{v1,vb,v3}
\fmf{plain}{v2,v3}
\fmffreeze
\fmf{wiggly}{va,vc,vb}
\fmfv{d.sh=circle,l.d=0, d.f=empty,d.si=.25w,l=$1$}{vc}
\fmfv{decor.shape=circle,decor.filled=30,decor.size=9}{v1}
\fmfv{decor.shape=circle,decor.filled=30,decor.size=9}{v2}
\fmfv{decor.shape=circle,decor.filled=30,decor.size=9}{v3}
\end{fmfgraph*}}=
P_3 (8 \omega -12)+P_6 \left(8 \omega +\frac{6}{\omega -2}\right)-\frac{4 P_4 (2 \omega -3) (3 \omega -5) (4 \omega -5)}{(\omega -2)^2}
\]

\[
\hspace{-7cm}
I_6=\int d^{2\o}x_1
\parbox{20mm}{
\begin{fmfgraph*}(50,50)
\fmftop{v1}
\fmfbottom{v2,v3}
\fmf{plain,right=0.75}{v1,va}
\fmf{dashes,right=0.75}{va,vb}
\fmf{plain}{vb,v2}
\fmf{plain,left=0.75}{v1,va}
\fmf{dashes,left=0.75}{va,vb}
\fmf{plain}{vb,v3}
\fmf{plain}{v2,v3}
\fmfv{decor.shape=circle,decor.filled=30,decor.size=9}{v1}
\fmfv{decor.shape=circle,decor.filled=30,decor.size=9}{v2}
\fmfv{decor.shape=circle,decor.filled=30,decor.size=9}{v3}
\end{fmfgraph*}}
=-2\,G^2(1,1)\,G(2,1)
\]

\[
\hspace{-2cm}
I_8=\int d^{2\o}x_1
\parbox{20mm}{
\begin{fmfgraph*}(50,50)
\fmftop{v1}
\fmfbottom{v2,v3}
\fmf{plain}{v1,va,vd,v2}
\fmf{plain}{v1,vb,vc,v3}
\fmf{plain}{v2,v3}
\fmffreeze
\fmf{wiggly}{va,vc}
\fmf{wiggly}{vb,vd}
\fmfv{decor.shape=circle,decor.filled=30,decor.size=9}{v1}
\fmfv{decor.shape=circle,decor.filled=30,decor.size=9}{v2}
\fmfv{decor.shape=circle,decor.filled=30,decor.size=9}{v3}
\end{fmfgraph*}}=
P_1 \left(-74 \omega -\frac{2 (\omega (5 \omega (5 \omega
  +17)-529)+532)}{(\omega -2)^2 (4 \omega -7)}\right)\]
\[+\frac{8 P_2 (2
  \omega -3) (\omega (9 \omega -25)+17)}{(\omega -2)^2}+P_3 \left(7
\omega +\frac{5}{\omega -2}+\frac{5}{28-16 \omega
}-\frac{7}{4}\right)\]
\[+2 P_4 \left(-504 \omega -\frac{1004}{\omega
  -2}-\frac{376}{(\omega -2)^2}-\frac{48}{(\omega -2)^3}+\frac{1}{3-2
  \omega }+\frac{50}{7-4 \omega }-181\right)\]
\[+\frac{P_5 (\omega -2) (2
  \omega -3)}{4 \omega -7}+\frac{2 P_6 (\omega -1) \omega (2 \omega
  -3)}{(\omega -2)^2}+P_7 \left(\frac{1}{2-\omega }-1\right)
\]

\[
\hspace{-3.5cm}
I_9=\int d^{2\o}x_1
\parbox{20mm}{
\begin{fmfgraph*}(50,50)
\fmftop{v1}
\fmfbottom{v2,v3}
\fmf{plain}{v1,va,v2}
\fmf{plain}{v1,vb,v3}
\fmf{plain}{v2,vd,vc,v3}
\fmffreeze
\fmf{wiggly}{va,vc}
\fmf{wiggly}{vb,vd}
\fmfv{decor.shape=circle,decor.filled=30,decor.size=9}{v1}
\fmfv{decor.shape=circle,decor.filled=30,decor.size=9}{v2}
\fmfv{decor.shape=circle,decor.filled=30,decor.size=9}{v3}
\end{fmfgraph*}}=
\frac{P_1 (3 \omega -4) (\omega (4 (\omega -7) \omega
  +57)-35)}{(\omega -2)^2 (4 \omega -7)}\]
\[+\frac{1}{8} P_3 \left(36
\omega +\frac{12}{\omega -2}+\frac{5}{4 \omega -7}-25\right)\]
\[+2 P_4
\left(174 \omega +\frac{328}{\omega -2}+\frac{118}{(\omega
  -2)^2}+\frac{12}{(\omega -2)^3}+\frac{1}{2 \omega -3}+\frac{25}{4
  \omega -7}+53\right)\]
\[+\frac{P_5 (\omega -2) (2 \omega -3)}{14-8
  \omega }-\frac{2 P_2 (2 \omega -3) (\omega (8 \omega
  -25)+19)}{(\omega -2)^2}
\]

\[
I_{10}=\int d^{2\o}x_1
\parbox{20mm}{
\begin{fmfgraph*}(50,50)
\fmftop{v1}
\fmfbottom{v2,v3}
\fmf{plain}{v1,va,v2}
\fmf{plain}{v1,vb,v3}
\fmf{plain}{v2,vc,v3}
\fmffreeze
\fmf{wiggly}{va,vb}
\fmf{wiggly}{va,vc}
\fmfv{decor.shape=circle,decor.filled=30,decor.size=9}{v1}
\fmfv{decor.shape=circle,decor.filled=30,decor.size=9}{v2}
\fmfv{decor.shape=circle,decor.filled=30,decor.size=9}{v3}
\end{fmfgraph*}}=
-\frac{P_1 (3 \omega -4) (\omega (2 \omega -5)+4)}{2 (\omega
  -2)^2}+P_2 \left(3 \omega +\frac{\omega (\omega +5)-12}{2 (\omega
  -2)^2}\right)\]
\[+P_3 \left(\frac{1}{2 (\omega
  -2)}+\frac{3}{4}\right)+P_4 \left(\frac{1}{2 \omega
  -3}-\frac{6}{\omega -2}-6\right)
\]

\[
\hspace{-4.5cm}
I_{11}=\int d^{2\o}x_1
\parbox{20mm}{
\begin{fmfgraph*}(50,50)
\fmftop{v1}
\fmfbottom{v2,v3}
\fmf{plain}{v1,va,v2}
\fmf{plain}{v1,vc,v3}
\fmf{plain}{v2,vb,v3}
\fmffreeze
\fmf{wiggly,tension=0.5}{va,vd}
\fmf{wiggly,tension=0.5}{vb,vd}
\fmf{wiggly,tension=0.5}{vc,vd}
\fmfv{decor.shape=circle,decor.filled=30,decor.size=9}{v1}
\fmfv{decor.shape=circle,decor.filled=30,decor.size=9}{v2}
\fmfv{decor.shape=circle,decor.filled=30,decor.size=9}{v3}
\end{fmfgraph*}}=
-\frac{2 P_4 (3 \omega -4) (4 \omega -5) (\omega -1)^2}{(\omega -2)^2
  (2 \omega -3)}\]
\[+\frac{P_2 (2 \omega -3) (\omega (4 \omega -13)+11)
  (\omega -1)}{(\omega -2)^3}+\frac{P_3 (2 \omega -3) (\omega -1)}{2
  (\omega -2)^2}\]
\[-\frac{P_1 (3 \omega -4) (2 (\omega -3) \omega +5)
  (\omega -1)}{(\omega -2)^3}
\]

\[
\hspace{-3.5cm}
I_{12}=\int d^{2\o}x_1
\parbox{20mm}{
\begin{fmfgraph*}(50,50)
\fmftop{v1}
\fmfbottom{v2,v3}
\fmf{plain}{v1,va,v2}
\fmf{plain}{v1,v3}
\fmf{plain}{v2,v3}
\fmfv{d.sh=circle,l.d=0, d.f=empty,d.si=.25w,l=$2$}{va}
\fmfv{decor.shape=circle,decor.filled=30,decor.size=9}{v1}
\fmfv{decor.shape=circle,decor.filled=30,decor.size=9}{v2}
\fmfv{decor.shape=circle,decor.filled=30,decor.size=9}{v3}
\end{fmfgraph*}}=\left(2\,G(1,5-2\o) + G(2,4-2\o)\right)Z_\text{scalar}
\]

\subsubsection{Two-point diagrams}

\[
T_1=-\frac{1}{4}T_7=~\,
\parbox{20mm}{
\begin{fmfgraph*}(50,50)
\fmfleft{v1}
\fmfright{v3}
\fmf{plain,left=0.3, tension=1}{v1,v4,v3,v2,v1}
\fmffixed{(0,33)}{v2,v4}
\fmf{wiggly,left=.3}{v2,v4,v2}
\fmfv{decor.shape=circle,decor.filled=30,decor.size=9}{v1}
\fmfv{decor.shape=circle,decor.filled=30,decor.size=9}{v3}
\end{fmfgraph*}}=
-\frac{1}{4}~~~
\parbox{20mm}{
\begin{fmfgraph*}(50,50)
\fmfleft{v1}
\fmfright{v3}
\fmf{plain,left=0.3, tension=1}{v1,v4,v3,v2,v1}
\fmffixed{(0,33)}{v2,v4}
\fmf{dashes,left=.3}{v2,v4,v2}
\fmfv{decor.shape=circle,decor.filled=30,decor.size=9}{v1}
\fmfv{decor.shape=circle,decor.filled=30,decor.size=9}{v3}
\end{fmfgraph*}}=
\frac{P_3 (3-2 \omega )}{4 (3 \omega -4)}\]
\[+\frac{1}{6} P_4
\left(\frac{1}{3 \omega -4}+\frac{3}{2 \omega -3}+10\right)
\]

\[T_2=~~\parbox{20mm}{
\begin{fmfgraph*}(50,50)
\fmfleft{v1}
\fmftop{t}
\fmfright{v2}
\fmfbottom{b1,b2}
\fmf{phantom,tension=2.95}{t,va}
\fmf{phantom,tension=2.15}{b1,vb1}
\fmf{phantom,tension=2.15}{b2,vb2}
\fmf{plain,left=0.3}{v1,va}
\fmf{plain,left=0.3}{va,v2}
\fmf{plain,right=0.25}{v1,vb1}
\fmf{plain,right=0.25}{vb1,vb2}
\fmf{plain,right=0.25}{vb2,v2}
\fmffreeze
\fmf{wiggly}{va,vb1}
\fmf{wiggly}{va,vb2}
\fmfv{decor.shape=circle,decor.filled=30,decor.size=9}{v1}
\fmfv{decor.shape=circle,decor.filled=30,decor.size=9}{v2}
\end{fmfgraph*}}
=
\frac{P_3 (9-6 \omega )}{16-12 \omega }+P_4 \left(\frac{1}{8-6 \omega
}+\frac{1}{3-2 \omega }-4\right)+P_1-\frac{P_2}{2}
\]

\[T_3=2T_4=~~
\parbox{20mm}{
\begin{fmfgraph*}(50,50)
\fmfleft{v1}
\fmftop{t1,t2,t3}
\fmfright{v2}
\fmfbottom{b}
\fmf{phantom,tension=1.0}{t1,va1}
\fmf{phantom,tension=1.1}{t2,va2}
\fmf{phantom,tension=1.0}{t3,va3}
\fmf{phantom,tension=2.95}{b,vb}
\fmf{plain,left=0.17}{v1,va1}
\fmf{plain,left=0.15}{va1,va2}
\fmf{plain,left=0.15}{va2,va3}
\fmf{plain,left=0.17}{va3,v2}
\fmf{plain,right=0.3}{v1,vb}
\fmf{plain,right=0.3}{vb,v2}
\fmffreeze
\fmf{wiggly,left=0.75}{va1,va3}
\fmf{wiggly}{va2,vb}
\fmfv{decor.shape=circle,decor.filled=30,decor.size=9}{v1}
\fmfv{decor.shape=circle,decor.filled=30,decor.size=9}{v2}
\end{fmfgraph*}}=
2~~~\parbox{20mm}{
\begin{fmfgraph*}(50,50)
\fmfleft{v1}
\fmftop{t1,t2,t3}
\fmfright{v2}
\fmfbottom{b}
\fmf{phantom,tension=1.0}{t1,va1}
\fmf{phantom,tension=1.1}{t2,va2}
\fmf{phantom,tension=1.0}{t3,va3}
\fmf{phantom,tension=2.95}{b,vb}
\fmf{plain,left=0.17}{v1,va1}
\fmf{plain,left=0.15}{va1,va2}
\fmf{plain,left=0.15}{va2,va3}
\fmf{plain,left=0.17}{va3,v2}
\fmf{plain,right=0.3}{v1,vb}
\fmf{plain,right=0.3}{vb,v2}
\fmffreeze
\fmf{wiggly}{va1,vcen}
\fmf{wiggly}{va3,vcen}
\fmf{wiggly}{vb,vcen}
\fmfv{decor.shape=circle,decor.filled=30,decor.size=9}{v1}
\fmfv{decor.shape=circle,decor.filled=30,decor.size=9}{v2}
\end{fmfgraph*}}=
P_1 \left(\frac{2}{2 \omega -3}+4\right)+\frac{P_3 (5-4 \omega )}{4-3
  \omega }\]
\[-\frac{2 P_4 (\omega -1) (4 \omega -5) (8 \omega -11)}{(3-2
  \omega )^2 (3 \omega -4)}-4 P_2
\]

\[T_5=~~
\parbox{20mm}{
\begin{fmfgraph*}(50,50)
\fmfleft{v1}
\fmfright{v3}
\fmf{plain,left=0.3, tension=1}{v1,v4,v3,v2,v1}
\fmffixed{(0,33)}{v2,v4}
\fmf{wiggly}{v2,vc,v4}
\fmfv{d.sh=circle,l.d=0, d.f=empty,d.si=.25w,l=$1$}{vc}
\fmfv{decor.shape=circle,decor.filled=30,decor.size=9}{v1}
\fmfv{decor.shape=circle,decor.filled=30,decor.size=9}{v3}
\end{fmfgraph*}}=~
\frac{P_3 (5-4 \omega )}{3 \omega -4}+P_4 \left(\frac{2}{12-9 \omega
}+\frac{12}{\omega -2}+\frac{40}{3}\right)
\]

\[T_6 = ~~
\parbox{20mm}{
\begin{fmfgraph*}(50,50)
\fmfleft{v1}
\fmfright{v3}
\fmf{plain,left=1., tension=1}{v1,va}
\fmf{dashes,left=1., tension=1}{va,vb}
\fmf{plain,left=1., tension=1}{vb,v3}
\fmf{plain,right=1., tension=1}{v1,va}
\fmf{dashes,right=1., tension=1}{va,vb}
\fmf{plain,right=1., tension=1}{vb,v3}
\fmfv{decor.shape=circle,decor.filled=30,decor.size=9}{v1}
\fmfv{decor.shape=circle,decor.filled=30,decor.size=9}{v3}
\end{fmfgraph*}}=
-G^3(1,1)
\]

\[T_8 = ~~
\parbox{20mm}{
\begin{fmfgraph*}(50,50)
\fmfleft{v1}
\fmftop{t1,t2}
\fmfright{v2}
\fmfbottom{b1,b2}
\fmf{phantom,tension=1.75}{t1,va1}
\fmf{phantom,tension=1.75}{t2,va2}
\fmf{phantom,tension=1.75}{b1,vb1}
\fmf{phantom,tension=1.75}{b2,vb2}
\fmf{plain,left=0.25}{v1,va1}
\fmf{plain,left=0.25}{va1,va2}
\fmf{plain,left=0.25}{va2,v2}
\fmf{plain,right=0.25}{v1,vb1}
\fmf{plain,right=0.25}{vb1,vb2}
\fmf{plain,right=0.25}{vb2,v2}
\fmffreeze
\fmf{wiggly}{va1,vb2}
\fmf{wiggly}{va2,vb1}
\fmfv{decor.shape=circle,decor.filled=30,decor.size=9}{v1}
\fmfv{decor.shape=circle,decor.filled=30,decor.size=9}{v2}
\end{fmfgraph*}}=~
P_1 \left(\frac{7}{(\omega -2)^2}+\frac{35}{4 (4 \omega
  -7)}+\frac{3}{3-2 \omega }+\frac{35}{2 (\omega
  -2)}+\frac{43}{4}\right)\]
\[+\frac{P_2 (4 (14-5 \omega ) \omega
  -38)}{(\omega -2)^2}+\frac{1}{24} P_3 \left(\frac{15}{4 \omega
  -7}+\frac{16}{4-3 \omega }-\frac{30}{\omega -2}-37\right)
-\frac{P_5 (\omega -2)}{4 (4 \omega -7)}\]
\[+P_4
\left(\frac{140}{(\omega -2)^2}+\frac{24}{(\omega -2)^3}+\frac{11}{2
  \omega -3}+\frac{50}{4 \omega -7}+\frac{3}{(3-2 \omega
  )^2}+\frac{4}{12-9 \omega }+\frac{222}{\omega
  -2}+\frac{401}{3}\right)\]

\[T_9 = ~~\parbox{20mm}{
\begin{fmfgraph*}(50,50)
\fmfleft{v1}
\fmfright{v3}
\fmf{plain,left=0.3, tension=1}{v1,v4,v3,v2,v1}
\fmffixed{(0,30)}{v2,v4}
\fmfv{d.sh=circle,l.d=0, d.f=empty,d.si=.25w,l=$2$}{v4}
\fmfv{decor.shape=circle,decor.filled=30,decor.size=9}{v1}
\fmfv{decor.shape=circle,decor.filled=30,decor.size=9}{v3}
\end{fmfgraph*}}=~~~
2\,G(1,4-2\o)\,Z_\text{scalar}
\]

\section{${\cal N}=4$ SYM details}
\label{app:n4}

We follow the conventions given in \cite{Erickson:2000af}. The
relevant terms in the Euclidean action arise from the scalar potential
and gauge coupling and are given by
\be
S = \int d^{2\o} x\left(f^{abc}\p_\m\Phi^{Ia}A^{b\m}\Phi^{Ic}
+\frac{1}{4} f^{abe}f^{cde}
\Phi^{Ia}\Phi^{Jb}\Phi^{Ic}\Phi^{Jd}\right) + \text{not relevant}.
\ee
We use Feynman gauge where the gauge field propagator is
\be
\la A^a_\m A^b_\n \ra = g_{YM}^2\frac{\d_{\m\n}\d^{ab}}{p^2}.
\ee
The one-loop correction to the scalar field is \cite{Erickson:2000af}
\be
\la \Phi^{Ia} \Phi^{Jb} \ra = g_{YM}^2 \frac{\d^{IJ}\d^{ab}}{p^2} 
-2 \,g_{YM}^4N \,G(1,1) \,\frac{\d^{IJ}\d^{ab}}{p^{6-2\o}}.
\ee
The integrals we need to evaluate for the ${\cal N}=4$ SYM example are
as follows (c.f. appendix J of \cite{Minahan:2009wg})
\[
\parbox{20mm}{
\hspace{-7.5cm}
\begin{fmfgraph*}(60,60)
\fmfleft{i}
\fmfright{o}
\fmftop{t}
\fmf{plain,tension=1.4}{i,v1}
\fmf{plain,tension=1.4}{v2,o}
\fmf{plain, right,tension=1}{v1,v2}
\fmffreeze
\fmf{phantom, tension=.8}{t,v3}
\fmf{dbl_plain, right=0.4,tension=0.4}{v3,v1}
\fmf{plain, left=0.64,tension=0.4}{v3,v1}
\fmf{plain, right=0.4,tension=0.4}{v2,v3}
\end{fmfgraph*}}
\hspace{-7.5cm}
~~= G(1,2)\,G(1,4-\o),\]
\vspace{-1.cm}
\[
\parbox{20mm}{
\hspace{-4cm}
\begin{fmfgraph*}(50,50)
\fmfleft{i}
\fmfright{o}
\fmf{plain, tension=2}{i,v1}
\fmf{plain, tension=2}{v3,o}
\fmf{plain,left=0.4, tension=1}{v1,v4,v3,v2,v1}
\fmffixed{(0,26)}{v2,v4}
\fmf{plain}{v2,v4}
\end{fmfgraph*}}
\hspace{-4cm}
= G(1,1)\left[\D(1,1)+2C(1,1)\,G(3-\o,2)\right],
\]
\vspace{-0.5cm}
\[
\parbox{20mm}{
\begin{fmfgraph*}(50,50)
\fmfleft{i}
\fmfright{o}
\fmf{plain, tension=2}{i,v1}
\fmf{plain, tension=2}{v3,o}
\fmf{plain,left=0.4, tension=1}{v1,v4,v3,v2,v1}
\fmffixed{(0,26)}{v2,v4}
\fmf{plain}{v2,v4}
\fmffreeze
\fmf{dbl_plain,left=0.4, tension=1}{v1,v4}
\end{fmfgraph*}}
=G(1,1)\left[\D(2,1)+C(2,1)\,G(3-\o,3)+C(1,2)\,G(4-\o,2)\right],
\]
where 
\be
C(\a,\b) = \frac{\a}{\a+\b+2-2\o},\quad
\D(\a,\b) = -\frac{\a\,G(\a+1,\b)+\b\,G(\a,\b+1)}{\a+\b+2-2\o},
\ee
and where $G(\a,\b)$ is given in (\ref{Gdef}).

\end{fmffile}
\bibliography{c3po}%
\end{document}